\documentclass[runningheads]{llncs}
\usepackage{graphicx}
\usepackage{listings}
\usepackage{xcolor}
\usepackage{tikz}
\usepackage{tkz-graph} 
\usetikzlibrary{shapes,arrows, automata}
\usetikzlibrary{positioning}
\usepackage{xargs}
\usepackage[colorinlistoftodos,prependcaption,textsize=tiny]{todonotes}
\usepackage{amsmath}
\usepackage{algorithm}
\usepackage{algpseudocode}

\lstdefinestyle{enhaut}{
  float=tp,
  floatplacement=tbp,
  abovecaptionskip=-5pt
}
\begin{document}
\title{Designing FSMs Specifications from Requirements 
with GPT 4.0\thanks{Supported by organisation x.}}
%
%
\author{Omer Nguena Timo\inst{1} 
\and Paul-Alexis Rodriguez\inst{2} 
\and Florent Avellaneda\inst{3}
}
\authorrunning{Omer Nguena Timo et al.}
\institute{Université du Québec en Outaouais, Quebec-Canada \\
\email{omer.nguena-timo@uqo.ca}\\
\and
Université Paris-Saclay, Paris-France\\
\email{paul-alexis.rodriguez@universite-paris-saclay.fr}\\
\and
Université du Québec à Montréal, Québec-Canada\\
\email{florent.avellaneda@uqam.ca}
}
\maketitle             
\begin{abstract}
Finite state machines (FSM) are executable formal specifications of reactive systems. These machines are designed based on systems’ requirements. The requirements are often recorded in textual documents written in natural languages. FSMs play a crucial role in different phases of the model-driven system engineering (MDE). For example, they serve to automate testing activities. FSM quality is critical: the lower the quality of FSM, the higher the number of faults surviving the testing phase and the higher the risk of failure of the systems in production, which could lead to catastrophic scenarios.  Therefore, this paper leverages recent advances in the domain of LLM to propose an LLM-based framework for designing FSMs from requirements. The framework also suggests an expert-centric approach based on FSM mutation and test generation for repairing the FSMs produced by LLMs. This paper also provides an experimental analysis and evaluation of LLM’s capacities in performing the tasks presented in the framework and FSM repair via various methods. The paper presents  experimental results with simulated data. These results and methods bring a new analysis and vision of LLMs that are useful for further development of machine learning technology and its applications to MDE.

\end{abstract}
\section{Introduction}
Designing formal specification of software systems is useful for the efficient analysis, code generation,  testing, verification, and maintenance of the systems. Formal specifications concerning the architecture or the behaviour of the systems are often designed from the system descriptions expressed in a natural language. Formal specifications are described with  models such as finite state automaton, Mealy machine, UML class diagram. This design task is time consuming, if entirely performed by humans. Its  automation can take advantage of recent advances in the field of machine learning and large language models (LLM).  

The class of state-transition models, allowing to specify system behaviour, have been widely used in model-based system engineering (MDE). Within this class, inputs that trigger transitions can be represented with identifiers or  complex Boolean expressions on variables, or both. The former representation is less detailed and suitable for high-level analyses. Mealy machines or finite state machines (FSM) are examples of less detailed models serving to study high-level system properties and to devise complex system analysis methods.  More detailed models involve accurate information. Examples of detailed models include UML state diagrams, symbolic automata, Simulink Stateflow diagrams, Verilog models and programs. Partial automation of the formal specification design can be considered, especially when the complete automation reduces to undecidable problems or produces approximate specifications. Partial automation involves domain-specific knowledge from human-experts to quickly make important decisions in the improving process for  approximate specifications.

Machine learning has served to automate specification design tasks and other tasks of the MDE~\cite{mao1998reusability}. It is now increasingly used ~\cite{oakes2024building,abukhalaf2024pathocl} due to recent progress with large language models (LLMs). A large language model (LLM)  is an artificial intelligence model designed to understand, generate, and manipulate human language. 
 Some LLMs include OpenAI's GPT series (like GPT-3 and GPT-4), Meta's LLaMA series, Google's BERT, and Microsoft's Turing-NLG.
The work in \cite{abukhalaf2024pathocl} uses GPT-4 to generate OCL constraints on attributes and associations in UML class diagrams, which specify software architecture. These constraints can be properties that high-quality UML diagrams must satisfy. The work in \cite{bhandari2024,liu2023verilogeval,thakur2024verigen} demonstrates the utility of LLMs in generating system behaviour specification or code and tests. In particular,~\cite{thakur2024verigen} uses LLMs to generate HDL programs from Verilog specifications. Verilog specifications are  state-transition machines. To evaluate the quality of the generated HDL program, \cite{thakur2024verigen} uses Verilog specification  created by the authors or borrowed from a known description repository\cite{HenryWong2019}; they also consider manually created and automatically generated test data for the generated HDL programs.~\cite{thakur2024verigen} does not study the automatic repair of faulty generated models.  Automatic repair of generated models is needed when the manual repair  is complex and time consuming.  This complexity depends on the type of faults observed in  generated models. Methods based on  distinguishing tests and checking sequences have been extensively investigated for creating test permitting to uncover faults in FSM specifications. To the best of our knowledge these methods have not been used in  LLM-based approach for designing specifications.

In this article, we use simulated data to investigate the design of finite state (Mealy) machines from their English descriptions. Our investigation aims to estimate the accuracy of GPT-4 in generating specification described with Mealy machines and to propose repair methods for faulty generated specification.  GPT-4.0 was selected because it provides a programming library  and it is used in other work. The simulated data is representative of real data and facilitates our investigation. Our contributions are as follows: 

\begin{itemize}
    \item We generate simulated data that we use to evaluate our approach. The simulated data include randomly generated  oracle FSMs, automatically generated requirements for the oracles FSMs.
    \item We device  LLM prompts for producing FSMs from from their textual descriptions
    \item We propose four approaches for repairing 
    produced FSMs in case they are faulty. The first approach is based on syntactic faults, and the other approaches consider erroneous behaviour detected with distinguishing or checking sequences. 
    \item We evaluate the capability of LLM to produce correct FSMs from their textual descriptions and we evaluate efforts from experts during the application of the semi-automated repairing method.   
\end{itemize}

The paper is organised as follows. First, we  introduce general concepts for  Mealy machines and the workflow of our investigations.  Then, we describe the production of simulated data for FSM and their descriptions, the generation of specification with GPT-4.0 and  repair methods for the generated specification. We report several experiments evaluating the proposed approaches. Finally, we summarise our main findings.

\section{LLM-based Approach for Designing FSM}
\newcommand{\calS}{\mathcal{S}}
\newcommand{\calM}{\mathcal{M}}
\newcommand{\X}{X}
\newcommand{\Y}{Y}
\newcommand{\x}{x}
\newcommand{\y}{y}

We consider the problem of automatically designing deterministic finite State Machine (DFSM) from their descriptions expressed in English. In other words, given a textual description of a machine, we want to automate the creation of a rigorous representation of the machine. A DFSM is a device responding to each input from its environment by producing an output and changing its internal state. It starts its execution from a known initial state.

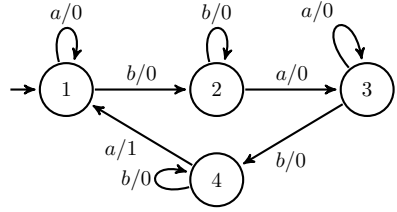
\begin{figure}[t]
    \centering
    \scalebox{0.80}{
\begin{tikzpicture}[->,>=stealth',shorten >=1pt,auto,node distance=2.5cm,initial text={},
every initial by arrow/.style={->}]
{\fontsize{10}{10}\selectfont
\tikzstyle{every node}=[scale=1,inner sep=1pt, outer sep=1pt,line width=1pt]
\tikzstyle{every edge}=[scale=1,line width=1pt, draw]
\tikzstyle{every state}=[fill=white,draw,text=black]
  \node[initial,state]  (S1)   at (0,5)                {$1$};
  \node[state]         (S2)  at ([xshift=2.5cm,yshift=0cm]S1) {$2$};
  \node[state]         (S3) at ([xshift=2.5cm,yshift=0cm]S2) {$3$};
  \node[state]         (S4) at ([xshift=0cm,yshift=-1.5cm]S2) {$4$};
  \path (S1) edge [loop above] node {$a/0$} (S1)
             edge          node[above] {$b/0$}(S2) 
        (S2) edge [loop above] node {$b/0$} (S2)
             edge  node[above, pos=0.5] {$a/0$} (S3)
        (S3)edge [loop, in=100, out=130,looseness=10] node {$a/0$ } (S3) 
             edge node[pos=0.7] {$b/0$ } (S4)
        (S4) edge [loop left]  node {$b/0$ } (S4)          
             edge node {$a/1$ } (S1) ;
}
\end{tikzpicture}
}
    \caption{A graphical rigorous  representation of a DFSM playing the role of an oracle}
    \label{fig:fsm:running:graph}
\end{figure}

A rigorous representation of  
a {\em finite state machine} (FSM) is a 5-tuple $\calS = (S, s^0, \X, \Y, T)$, where $S$ is a finite set of states with initial state $s^0$; $X$ and $Y$ are finite non-empty disjoint sets of inputs and outputs, respectively; $T \subseteq S \times \X \times \Y \times S$ is a transition relation and a tuple $(s, x, y, s') \in T$ is called a transition from $s$ to $s'$ with input $x$ and output $y$. The set of transitions from state $s$ is denoted by $T(s)$. $T(s,x)$ denotes the set of transitions in $T(s)$ with input $x$.   For a transition $t = (s, x, y, s')$, we define $src(t) = s$, $inp(t)=x$, $out(t)= y$ and $tgt(t) =s'$.
The set of uncertain transitions in an object $A$ is denoted by $Unctn(A)$.
Transition $t$ is {\em uncertain} if $|T(src(t), inp(t))| > 1$, i.e., several transitions  from the $src(t)$ have the same input as $t$; otherwise $t$ is {\em certain}. We say that $\calS$ is \textit{deterministic} (DFSM) if it has no uncertain transition, otherwise it is non-deterministic (NFSM). $\calS$ is {\em completely specified} (input complete FSM) if for each tuple $(s, x) \in  S \times \X$ there exists transition $(s, x, y, s') \in T$. A rigorous description can be coded in a rigorous language such as the DOT language \cite{gansner2006drawing}. 
Figure~\ref{fig:fsm:running:graph} presents a DFSM composed of $4$ states and $8$ transitions. 
\begin{lstlisting} [language={},
numbers=none,
stringstyle=\ttfamily,
    showstringspaces=false,
    breaklines=true,
    frameround=ffff,
    frame=single,
    backgroundcolor = \color{green}, style=enhaut, label=lst:fsm:running:engl:1, caption= A description of DFSM in Figure~\ref{fig:fsm:running:graph}]
DFSM_description = "when it is in state s1 , 0 is returned and the application moves to state s2 on occurence of b. in state s1 it returns 0 and it moves to state s1 if the input a occurs., 0 is returned and the application reaches state s3 on occurence of input a in  state s2., 0 is returned and it reaches s2 if  b occurs in  state s2. from state s3 , 0 is produced and the system reaches s4 on occurence of input b.
 the output 0 is produced and the system moves to s3 if the input is a when the system is in state s3. the application produces 1 and it reaches s1 on occurence of input a from state s4. when the system is in state s4 , 0 is returned and state s4 is reached on occurence of b. state s1 is the initial state."

\end{lstlisting}

\par An {\em execution of $\calS$ in $s$}, $e= t_1t_2\ldots t_n$ is a finite sequence  of transitions forming a path from $s$ in the state transition diagram of $\calS$, i.e., $src(t_1)=s$, $src(t_{i+1}) = tgt(t_i)$ for every $i=1...n-1$.  
A trace $\x/\y$ is a pair of an input sequence $\x$ and an output sequence $\y$, both of the same length. The trace of $e$ is $inp(t_1)inp(t_2)\ldots inp(t_n)/ out(t_1)out(t_2)\ldots out(t_n)$.
A trace of $\calS$  in $s$ is a trace of an execution of $\calS$ in $s$. Let $Tr_\calS(s)$ denote the set of all traces of $\calS$ in  $s$. $Tr_\calS$ denotes the
set of traces of $\calS$ in the initial state $s^0$ and it represent the behaviour of $\calS$. Given a sequence $\beta \in (\X\Y)^*$, the input (resp. output)
projection of $\beta$, denoted $\beta_{\downarrow X}$ (resp. $\beta_{\downarrow \Y}$), is a sequence obtained
from $\beta$ by erasing symbols in $Y$  (resp. $\X$); if $\beta$ is the trace of execution $e$, then $\beta_{\downarrow X} = inp(e)$ (resp. $\beta_{\downarrow \Y} = out(e)$) is called the input (resp. output) sequence of $e$ and we say that $out(e)$ is the \textit{response} of $\calS$ in $s$  to (the application of) input sequence $inp(e)$. $|X|$ denotes the size of set $X$.

A textual description of a DFSM corresponds to english sentences that describe the DFSM. The description can contain structural information on the identification of states and transitions. It can also  contain behaviour information associating input sequences to the corresponding output sequences. A DFSM corresponds to many descriptions. 
\begin{figure}[t]
    \centering
    \includegraphics[scale=0.35]{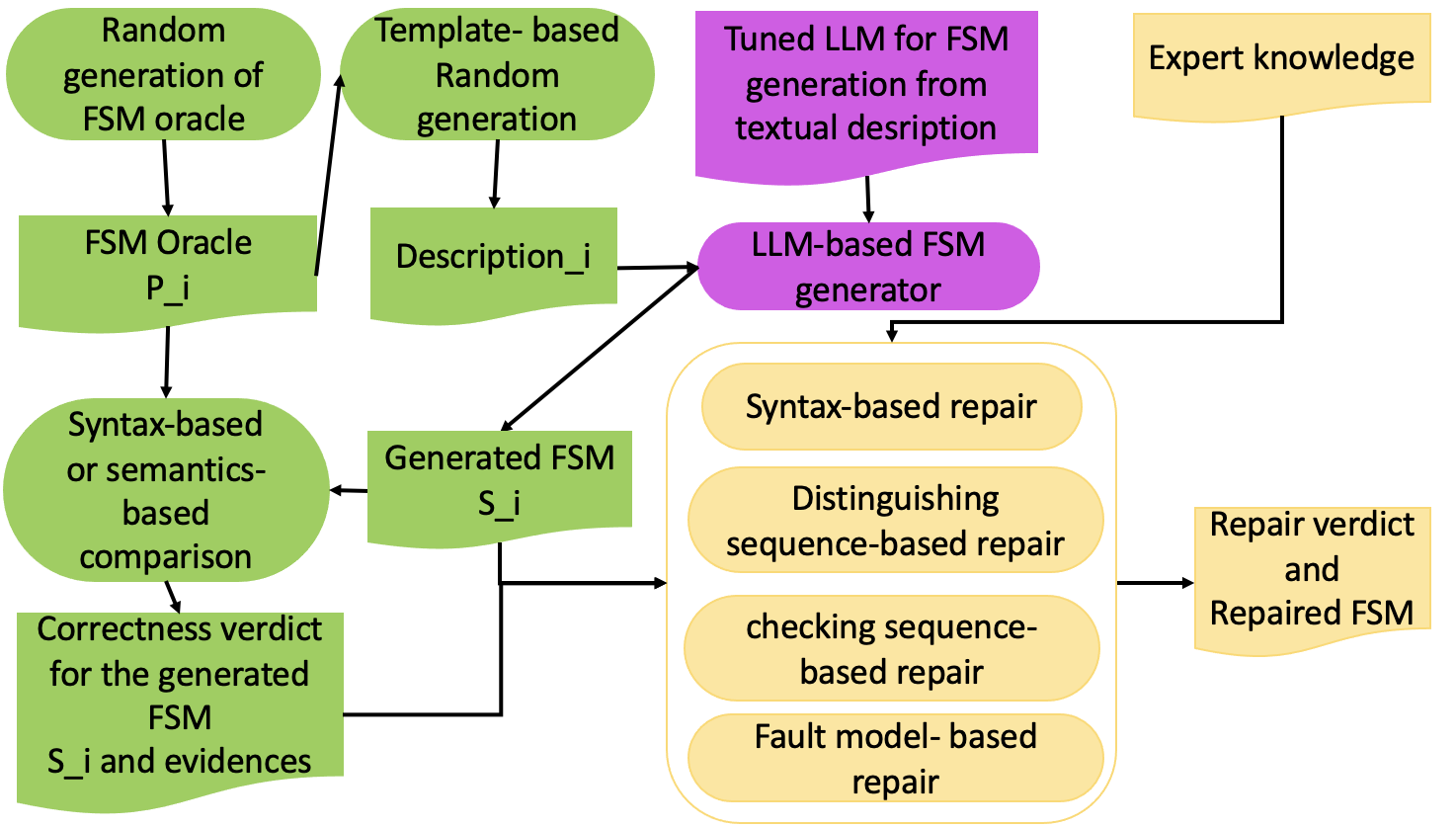}
    \caption{LLM-Based approach to design FSMs from their textual descriptions}
    \label{fig:framework}
\end{figure}
Listing~\ref{lst:fsm:running:engl:1} presents a description for the DFSM oracle in Figure~\ref{fig:fsm:running:graph}.  The textual descriptions were synthetically generated. They do not correspond to the description of a real machine. Note that descriptions do not always introduce keywords (state, input, output) before the  identifiers of states, inputs or outputs. It is more challenging for LLMs to analyse synthetic description because the domain-specific context is missing.

Our approach in Figure~\ref{fig:framework} to design DFSM from their description leverages capabilities of large language models (LLMs). It includes prompt-engineering of LLMs to generate DFSMs from their descriptions. It also include repair methods for faulty generated DFSMs. These methods rely on information from human-experts. Only human experts playing the role of oracles can provide the needed information.   
\section{GPT-based Generation of DFSMs from  Descriptions}
We exploit the prompting features offered by GPT LLM models. Prompting consists in interacting with LLMs by setting parameters and providing them with specific queries (prompts) to generate desired answers (outcomes). We use GPT-4 since it demonstrated the best promising results. 

Elaborating prompts, also known as prompt engineering~\cite{denny2023conversing,white2023prompt}, is a manual activity left to the expertise of engineers. There is no formal method describing how to achieve this task.  It  often starts by a \emph{fine-tuning} activity where  queries and expected answers are given to the LLMs for them to better understand the context of task they are expected to achieve. The \emph{fine-tuning} process can also be performed by defining a \textit{system role} in the message addressed to the LLM. The system role can add specific context to LLMs and specify the task we want to achieve, as well as examples of queries and expected outcomes. 
Listing~\ref{lst:prompt:finetuning:query} presents the Python code we use for fine-tuning and querying  GPT-4o. The  \textit{user role} of the message transmitted to LLM allow specifying the query. The message contains the textual description in Listing~\ref{lst:fsm:running:engl:1}. %

\begin{lstlisting}[
numbers=none,
stringstyle=\ttfamily,
    showstringspaces=false,
    breaklines=true,
    frameround=ffff,
    frame=single,
    backgroundcolor = \color{pink}, style=enhaut, label=lst:prompt:finetuning:query, caption=Prompt given to GTP-4o for the generation of DFSMs, language=python]
machine_role = "You are a professional software engineer working on a project to generate a CSV representation of a finite state machine (FSM) from a natural language description. You have been given the following description:"

prompt = f"{DFSM_description} Can you create the previous automaton on csv format with the following order: State, Input, Output, Next_State, the states should be named Si (where i is always a number), the first row should contain State, Input, Output, Next_State, and the other rows should only contain the state name in Si format (where i is always a number) the input the output and the next state name in Si format (where i is always a number), there shouldnt spaces between each information only comas., here is an example: first row: State, Input, Output, Next_State, second row: S0,a,0,S2 third row: S1,b,1,S3 fourth row: S2,c,0,S1 fifth row: S3,d,1,S0. Please do not add any comments to the csv file. Please keep in mind the machine should be complete and deterministic."
 
response = client.chat.completions.create(
        model="gpt-4o",
        temperature=0.0,
        messages=[
            {"role": "system", "content": machine_role},
            {"role": "user", "content": prompt}])
\end{lstlisting}

Note that LLM models can generate syntactic and semantics different DFSMs for different textual descriptions of the same machine. The syntactic difference may concern the name of the states, the number of states and  the transitions. The semantic difference concern the sets of input-output sequences of the generated DFSMs. Evaluating the correctness of generated DFSMs, as well as the capability of LLMs to achieve the generation task is desired. This evaluation is challenging because the result also depends on the fine-tuning activity of which the quality can not be formally guaranteed. The challenge remains, even under the assumption of the best fine-tuning. In the context of randomly generated DFSMs, fine tuning the model is not necessary. Fine-tuning the model allows for the LLM to better perform the desired task based on the new information given to it, however when working with randomly generated DFSMs there is no patterns or expected behaviours that can improve the LLM's performance in accomplishing the desired task. Nevertheless fine-tuning the prompt is critical and further supports the difficulties of evaluating the LLMs performances.  In the next section, we present evaluation methods and their empirical evaluations.

\section{Evaluating the Correctness of Generated DFSMs}
Correct generated DFSMs formally specify the expected behaviours in the textual description. Only oracles can verify the correctness of generated DFSMs. In practice, oracle are human experts. However, for automating the evaluation of the correctness of the generated DFSM, we perform a simulation experiment. In said experiment we mimic human experts with DFSMs playing the role of oracles. Such DFSMs are called DFSM oracles.

\subsection{Evaluation with  Oracles}
The evaluation approach consists in comparing the oracles with the generated DFSMs. We consider a comparison method based on the syntax and a comparison method based on the semantics of the DFSMs. The results of the syntax-based comparisons are set of transitions. The results for the semantics-based comparisons are distinguishing input sequences. These results are evidences of differences between DFSMs. Experiments with multiples oracles permit a statistic evaluation of the LLMs capability to generate DFSMs from their textual descriptions. The syntax-based comparison  of DFSMs considers the states and the transitions in  DFSMs. Such a comparison is possible if there is a  one-to-one correspondence between the states of the DFSMs. This correspondence can be  simple with the names of the states or complex with an isomorphism between the DFSMs. Simple correspondences is adequate for our work with synthetic data. Accordingly, two DFSMs syntactically differ if the following conditions hold:
\begin{itemize}
    \item They have different number of states or they use different set of states.
    \item They use the same set of states and at least one of the following fault conditions~\cite{ghedamsi1993multiple} hold:
    \begin{itemize} 
        \item (fault type 1: additional transition fault):  for some $(s,x,y,t)$ in $M_2$ there is no transition in $M_1$ from $s$ with input $x$
        \item (fault type 2: missing  transition fault): for some $(s,x,y,t)$ in $M_1$ there is no transition in $M_2$ from $s$ with input $x$ 
        \item (fault type 3: local output fault) there are  $(s,x,y,t)$ in $M_1$ and  $(s,x,y',t')$ in $M_2$ such that $y\neq y'$
        \item (fault type 4: transfer fault) there are  $(s,x,y,t)$ in $M_1$ and  $(s,x,y',t')$ in $M_2$ such that $t\neq t'$
    \end{itemize}
    \end{itemize}
There is a linear time algorithm for checking the syntactic difference between two DFSMs. Such an algorithm checks the conditions in the definition of the syntactic difference in each and every pair of states in $S_1\times S_2$ and for each and every input in $X$. The result of such an algorithm is a collection of identified faults represented by a set of transitions $\Delta$. Each transition  in $\Delta$ is  an evidence of a fault condition. Assuming that one of two DFSM plays the role of the expected DFSM, $\Delta$ can be partitioned in two transitions classes:   desired transitions or undesired transitions. \emph{Desired transitions} in $\Delta_d \subseteq \Delta$ occurs in the expected DFSM but they are missing in the other DFSMs. \emph{Undesired transitions}  in $\Delta_u \subseteq \Delta$ are missing in the expected DFSM whereas they occur in the other DFSM. The transition classes can serve to repair the generated DFSM.

\par The semantic-based comparison for two DFSMs considers their traces. Intuitively two DFSMs are not equivalent (i.e., different) if they have distinct set of traces. For input-complete DFSMs, the difference is observed if an input sequence corresponds to different output sequences in the DFSMs. 
Formally, given input sequence $\x \in \X^*$, let $out_\calS(s, \x)$ denote the set of output sequences which can be produced by $\calS$  when $\x$ is applied at state $s$, that is  $out_\calS(s, \x) = \{\beta_{\downarrow Y} \mid \beta\in Tr_\calS(s) \text{ and }\beta_{\downarrow X}= \x\}$.
Given state $s_1$ and $s_2$ of an FSM $\mathcal{S}$ and an input sequence $\x\in X^*$, $s_1$ and $s_2$ are $\x$-distinguishable, denoted by $s_1\not \simeq_\x s_2$ if  $out_\mathcal{S}(s_1,\x) \neq out_\mathcal{S}(s_2,\x)$; then $\x$ is called a distinguishing input sequence for $s_1$ and $s_2$. $s_1$ and $s_2$ are $\x$-equivalent, denoted by $s_1 \simeq_\x s_2$ if  $out_\mathcal{S}(s_1,\x) = out_\mathcal{S}(s_2,\x)$. $s_1$ and $s_2$ are distinguishable, denoted by $s_1\not \simeq s_2$,  if they are $\x$-distinguishable  for some input sequence $\x\in X^*$; otherwise they are equivalent. Let $a\in X$. A distinguishing input sequence $\x a \in X^+$ for $s_1$ and $s_2$ is {\em minimal} if $\x$ is not distinguishing for $s_1$ and $s_2$. 
Two complete DFSMs $\calS_1 = (S_1, s_1^0, X, Y, T_1)$ and $\calS_2 = (S_2, s_2^0, X, Y, T_2)$ over the same input and output alphabets are distinguished with input sequence $\x$ if $s_1^0\not \simeq_\x s_2^0$.

A linear time algorithm for performing the semantic-based comparison of two DFSMs can be formalised with the notion of distinguishing automaton. Such an automaton mimics parallel executions of the two DFSMs and identifies parallels executions that produced different output sequences for the same input sequences. Formally,  given an FSM ${\calS}_1 = (S_1, s_{1_0}, \X, \Y, T_1)$ and an FSM ${\calS}_2 = (S_2, s_{2_0}, \X, \Y, T_2)$ a finite automaton ${\cal D} = (D \cup \{\nabla\}, d_0, G, \Theta, \nabla)$, where $D \subseteq S_1 \times S_2$, $\nabla$ is an accepting (sink) state and $\Theta\subseteq D \times \X \times D $ is the transition relation is the {\em distinguishing automaton} for ${\calS}_1$ and ${\calS}_2$, if it holds that
\begin{itemize}
\item $d_0 =(s_{1_0},s_{2_0})$ is the initial state in $D$
\item For any $(s_1,s_2) \in D$
\begin{itemize}
\item $((s_1, s_2), \x, (s_1', s_2')) \in \Theta$, if there exists $(s_1, \x, \y, s_1') \in T_1$, $(s_2, \x, \y, s_2') \in T_2$
\item $((s_1, s_2), \x,  \nabla) \in \Theta$, if there exists $(s_1, \x, \y_1, s_1') \in T_1$, $(s_2, \x, \y_2, s_2') \in T_2$ and $y_1\neq y_2$. 
\end{itemize}
\end{itemize}
The language of ${\cal D}$, $L_{\cal D}$ is the set of  input sequences labelling executions of ${\cal D}$ from $d_0$ to the sink state $\nabla$. 
We can show that $\calS_1$ and $\calS_2$ are equivalent if and only if $L_{\cal D}=\emptyset$; otherwise input sequences in $L_{\cal D}$ are distinguishing for $\calS_1$ and $\calS_1$. Searching a distinguishing input sequence for $\calS_1$ and $\calS_1$ amounts to searching a path from $d_0$ to  $\nabla$ in ${\cal D}$; this can be done with a linear time algorithm. 
\subsection{Experimental Results}
We conduct an empirical evaluation of our approach to generate DFSM from their description in natural language. The experiment is based on a prototype tool we wrote in Python. The tool uses API provided for prompting GPT-4 and implements both syntactic-based and semantic-based comparison methods for DFSMs. The evaluation considers randomly generated description.

We devised a Python module that randomly creates expected DFSMs that play the role of the oracle in evaluating the correctness of DFSMs generated from descriptions with LLMs.  The module takes as inputs the number of state of the FSM, the sizes of the input and the output alphabets. DFSM  in Figure~\ref{fig:fsm:running:graph} was created by the module.
We also devised a Python module for generating FSM description. A description of an FSM is a collection of english sentences, one sentence per transition in the FSM.  We devised english \textit{sentence patterns} for describing a transition. Each transition description generation randomly selects a transition description pattern from a set of available patterns. Each transition description pattern  corresponds to a combination of patterns for describing the transition components (starting state, target state, input and output).  The description in Listing~\ref{lst:fsm:running:engl:1} is produced with the implemented patterns. 
These patterns are implemented with function \texttt{toNL}  in Listing~\ref{listing:pattern:state}, Listing~\ref{listing:pattern:transition} and Listing~\ref{listing:pattern:fsm} in Appendix.

GPT-4o produces a correct DFSM from the description in Listing~\ref{lst:fsm:running:engl:1}. 
The empirical evaluation considers DFSMs having $5$, $10$ and $25$ states. The DFSMs defines $5$ inputs and $2$ outputs. Domain-specific finite state machines often have less than 25 states. The higher the number of inputs and outputs, the higher the risk of wrong analysis of the description by LLMs. For a given number of states, we randomly generate $30$  oracles (i.e., expected DFSM). For each oracle, we produce a random description. For each description, we use GPT-4.0 to generate a DFSM. We then compare the generated DFSM and the oracle. The generated DFSM is called faulty if it differs from the oracle. Table~\ref{table:approach:syntax:result}, Table~\ref{table:approach:distseq:result}, Table~\ref{table:approach:checkseq:result}
present the number of faulty generated DFSMs with three different experiments. Each experiment serves also to evaluate a repair method. We observed that the most frequent faults made by GPT-4o are output faults and the missing transition faults. This observation can serve elaborating  fault models for repairing generated DFSMs. Table~\ref{table:evaluation:faults:mean:max} shows the mean numbers and the maximum numbers of fault types  per machine results for the following fault types: type 1 (mainly non-determinism in generated machines), type 2 (mainly missing transition on inputs),  type 3 and type 4.  For each number of states, we consider 30 oracles to produce the results.
\begin{table}[t]
    \centering
    \begin{tabular}{|c|c|c|c|c|c|} \hline 
      {\bf nb of states}  & {\bf Type 1} & {\bf Type 2} & {\bf Type 3} & {\bf Type 4} &{\bf All}\\ \hline \hline 
      \multicolumn{6}{|c|}{Means} \\ \hline 
        5 & 0.0 & 0.03 & 0.0 & 0.0 & {\bf 0.03}\\ \hline 
        10 & 0.07 & 0.43 & 0.27 & 0.33 & {\bf 1.1}\\ \hline \hline 
        \multicolumn{6}{|c|}{Maximum} \\ \hline
         5 & 0 & 1 & 0 & 0 & {\bf 1}\\ \hline 
        10 & 1 & 3 & 3 & 4 & {\bf 11}\\ \hline 
    \end{tabular}
    \caption{Mean and maximum numbers of fault types per machine in the generated DFSMs}
 \label{table:evaluation:faults:mean:max}
\end{table}

\section{Repairing Generated DFSMs}
This section addresses two interdependent challenges. The first challenge is the DFSM repair itself. The second is evaluating the capability of LLMs to understand and apply repair queries (prompts). The following subsections present four approaches for overcoming the challenges.

\subsection{Approach based on Identified Syntactic Faults with a DFSM Oracle\label{subsec:repair:approach:syntax}}

This approach assumes the existence of a DFSM oracle, which is unrealistic but allows the accomplishment of the second goal.
The approach works as follows. Let $\mathcal{P}$ be a DFSM oracle,  $\pi_{org}$ be the original fine-tuned prompt built from a description of $\mathcal{P}$ and sent to the LLM which returns the generated DFSM $\mathcal{S}_0$. If $\mathcal{S}_0$ syntactically differ from $\mathcal{P}$, we can construct a set $\Delta_0= \Delta_{u_0} \cup \Delta_{d_0} = \{\tau_1,\tau_2,\cdots, \tau_n\}$ of syntactic faults, wrong transitions, according to the errors mentioned in section 4.1, by performing a syntactic comparison between $\mathcal{P}$ and $\mathcal{S}_0$. If $\Delta_0$ is non-empty, then the generated machine is not correct. Therefore, we start the repair of it. For reparing the machine, each transition in $\Delta_i$ ($\Delta_0$ for the first comparison between $\mathcal{P}$ and $\mathcal{S}_0$, and $\Delta_i$ for the repair attempt number i where we compare $\mathcal{P}$ and $\mathcal{S}_i$) is added to the prompt $\pi_{i-1}$, produced by the previous repair attempt, to produce the enhanced prompt $\pi_i$. With $\pi_0 = \pi_{org}$. To produce $\pi_i$, we use a repair template. Each desired transition $\tau_d\in \Delta_{d_i}$ corresponds to an adding query  $\pi_{\tau_d}$ of  $\tau_d$ in $\mathcal{S}_i$, where $\pi_{\tau_d}$ is the corresponding sentence for desired transitions in the template for transition $\tau_d$. Each undesired transition $\tau_{u} \in \Delta_{u_i}$ corresponds to an erasing query $\pi_{\tau_{u}}$ of $\tau_{u}$ from $\mathcal{S}_i$, where $\pi_{\tau_{u}}$ is the corresponding sentence for undesired transitions in the template for transition $\tau_{u}$. Each correctly generated transition $\tau_c \in S_i$ corresponds to a conserving query $\pi_{\tau_c}$, where $\pi_{\tau_{c}}$ is the corresponding sentence for correctly generated transitions in the template for transition $\tau_{c}$. We consider $\pi_{\Delta_i} = \{\pi_{\tau_{u_1}},\cdots, \pi_{\tau_{u_n}},\pi_{\tau_{d_1}}, \cdots, \pi_{\tau_{d_n}}, \pi_{\tau_{c_1}}, \cdots, \pi_{\tau_{c_n}}  \}$. The concatenation of each member of $\pi_{\Delta_i}$ is called $\pi_{f}$, with $\pi_{f} = \pi_{\tau_{u_1}}\ldots \pi_{\tau_{u_n}}\pi_{\tau_{d_1}}\ldots \pi_{\tau_{d_n}}\pi_{\tau_{c_1}}\ldots \pi_{\tau_{c_n}}$. $\pi_{f}$ is the result of the template applied to the set $\Delta_i$. We then concatenate $\pi_{i-1}$ and $\pi_{f}$ to produce the improved prompt $\pi_i$ for the repair attempt number i. Then, this prompt is given to the LLM as a query, which will produce the DFSM $\mathcal{S}_i$. We repeat the process until the LLM produces a DFSM equivalent to the oracle or the maximum number of repair attempts has been reached. For each method the maximum number of repair attempts is calculated as follows: $max\_iter = |S|*|A|$ with S the set of states of the oracle and A the set of inputs for the oracle. The same calculation method is used in every method. Listing~\ref{lst:prompt:repairquery} presents the prompt for the syntatic-based repair.

\begin{lstlisting}[
numbers=none,
 stringstyle=\ttfamily,
    showstringspaces=false,
    breaklines=true,
    frameround=ffff,
    frame=single,
    backgroundcolor = \color{pink},style=enhaut, label=lst:prompt:repairquery, caption= Template for repair queries based on syntactic faults]
new_prompt = old_prompt
new_prompt += "Correct the automaton so that these transitions are present in the generated automaton:\n"
for tu in delta_u:
    new_prompt += tu.toNL() + "\n"
new_prompt += "Correct the automaton so that these transitions are not present in the generated automaton:"
for td in delta_d:
    new_prompt += td.toNL() + "\n"
new_prompt += "These transitions are correct and should be present in the generated automaton:\n"
for tc in g:
    new_prompt += tc.toNL() + "\n"
new_prompt += "Please keep this format: State, Input, Output, Next_State, the states should be named Si (where i is always a number), the first row should contain State, Input, Output, Next_State, and the other rows should only contain the state name in Si format (where i is always a number) the input the output and the next state name in Si format (where i is always a number), there shouldnt spaces between each information only comas., here is an example: first row: State, Input, Output, Next_State, second row: S0,a,0,S2 third row: S1,b,1,S3 fourth row: S2,c,0,S1 fifth row: S3,d,1,S0. Do not add any comments\n"
return new_prompt
\end{lstlisting}

Table~\ref{table:approach:syntax:result} presents the evaluation results of the effectiveness of the repair approach with randomly generated DFSM oracle. For each faulty generated DFSM, we estimate the size of $\Delta$. For 5 states the 2 faulty machines were corrected in 1 attempt. For 10 states 7 machines were corrected in 1 attempt, 1 machine in 2 attempts and 1 machine in 3 attempts. All were corrected. The success of this repair approach can be explained by the fact that feeding the LLM with specific syntactic changes to perform in the DFSM do not require any analysis of the behaviours of the DFSM. 

\begin{table}[t]
    \centering
\begin{tabular}{|c|p{1,5cm}|p{1,5cm}|p{1,2cm}|p{1,2cm}|p{1,5cm}|} \hline 
   {\bf nb states}  & nb of oracles & nb. of  faulty generated DFSM & average size of $\Delta$ & Max size of $\Delta$ & repair succeeding rate \\ \hline \hline 
    5 & 30 & 2 & 0.033 & 1 & 100\% \\ \hline 
    10 & 30 & 9 & 1.1 & 11 & 100\% \\ \hline
    25 & 1& 0 & 0 & 0 & 100\% \\ \hline
\end{tabular}
    \caption{Evaluation results   of the repair approach based on Syntactic faults}
    \label{table:approach:syntax:result}
\end{table}
\subsection{Approach based on Distinguishing  Sequences}

This approach is based on input-output traces and serve to evaluate the capability of LLMs to process prompts based on input-output traces. Traces define the semantics of DFSM. A strong assumption is that the expected DFSM or the DFSM oracle is available for the automatic computation of distinguishing sequences. 

The difference between this approach and the one presented in Section~\ref{subsec:repair:approach:syntax} is that repair queries are built using a template based on distinguishing sequences. Otherwise, the process is similar, and the iterative process is maintained. Given an oracle $\mathcal{P}$  and a generated DFSM $\mathcal{S}_i$, we use the distinguishing automata to estimate a subset $Iseq = \{\sigma_1,\sigma_2,\ldots,\sigma_n\}$ of the distinguishing input sequences for $\mathcal{P}$ and $\mathcal{S}_i$. $Iseq$ is empty if the two machines are equivalent. We estimate the sequences with a random search of accepting paths in the distinguishing automata. Defining other search strategies for this task is possible. Each input sequence $\sigma_i$ corresponds to an output sequence $\gamma_i$ in $\mathcal{P}$. For the trace $\tau_i = \sigma_i/\gamma_i$ we devise a repair query $\pi_{\tau_i}$. $\pi_{\tau_i}$ corresponds to the sentence in the template for the trace $\tau_i$, where we specify the expected output in $\mathcal{P}$ for the given input. Prompt $\pi_i$ in this method is built similarly to the previous method, with a difference in the contents of the $\pi_f$  that will be concatenated to $\pi_{i-1}$. In this method, $\pi_f$ corresponds to the concatenation of each $\pi_{\tau_i}$. $\pi_i$ is built as it was in the previous method. However, there is no warranty that this approach will terminate and enable the LLM to produce a specification consistent with the distinguishing sequences. There is a possibility that the LLM could loop in a specification space that does not contain the one expected by the oracle. The stop conditions and iterative process remain the same as previously.
 
Listing~\ref{lst:prompt:repairquery:distseq} presents the template for repair queries.
 \begin{lstlisting}[
 numbers=none,
 stringstyle=\ttfamily,
    showstringspaces=false,
    breaklines=true,
    frameround=ffff,
    frame=single,
    backgroundcolor = \color{pink}, style=enhaut,label=lst:prompt:repairquery:distseq, caption=Template for repair queries based on  distinguishing sequences]
    new_prompt = old_prompt
    new_prompt += "Correct the automaton so that this input sequence given to the automaton:\n"
    new_prompt += sigma_i + "\n"
    new_prompt += "Generates this output sequence:\n"
    new_prompt += gamma_i + "\n"
    new_prompt += "Please keep this format: State, Input, Output, Next_State, the states should be named Si (where i is always a number), the first row should contain State, Input, Output, Next_State, and the other rows should only contain the state name in Si format (where i is always a number) the input the output and the next state name in Si format (where i is always a number), there shouldnt spaces between each information only comas., here is an example: first row: State, Input, Output, Next_State, second row: S0,a,0,S2 third row: S1,b,1,S3 fourth row: S2,c,0,S1 fifth row: S3,d,1,S0.Do not add any comments\n"
    return new_prompt
\end{lstlisting}

Table~\ref{table:approach:distseq:result} presents the evaluation results  of the effectiveness of the repair approach with randomly generated DFSM oracle. For each faulty generated DFSM, we estimate the repair succeeding rate that corresponds to the ratio of correct DFSM repair candidate over the number of faulty generated DFSM (the corrected DFSM was corrected in 1 attempt). The results  are slightly worst than the previous syntax-based method. This can be explained by the mechanisms used in LLMs, specifically attention. When treating small amounts of information, the attention mechanism keeps a higher percentage of each input word to encode each word in the phrase, but when treating a higher amount of information, the mechanism will use a lesser portion of it. Checking sequences ore often longer than distinguishing sequences and introduce a high amount of information on the branching structure of the DFSMs. 

\begin{table}[t]
    \centering
\begin{tabular}{|c|c|p{2cm}|p{2cm}|} \hline 
   {\bf nb states}  & nb of oracles & nb. of  faulty generated DFSM & repair succeeding rate \\ \hline \hline 
    5 & 30 & 1 & 100\%\\ \hline 
    10 & 30 & 7 & 0\% \\ \hline 
\end{tabular}
    \caption{Evaluation results of the repair approach based on distinguishing input-output sequences}
    \label{table:approach:distseq:result}
\end{table}

\subsection{Approach based on Checking Sequence}
Realistic concerns motivate this approach. In practice, human experts play the role of oracles who can only infer the desired behaviour (input-output trace) of the system from their textual description. They do not know the whole structure (states and transitions)  of the expected DFSM. Even if this is true, we still want to evaluate the correctness of LLM-generated DFSM and correct them if faulty. We suggest an approach based on the so-called \textit{checking sequence}. A checking sequence for a  DFSM $S$ with $n$ states is a trace of $S$ that cannot be a trace of any other DFSM with $n$ states unless the latter is equivalent to $S$. Estimating the checking sequence for a DFSM is not trivial. Existing algorithms~\cite{lee1996principles,petrenko2019fsm} are NP-complete. In this method two options are available: whether we can use the oracle as the expert or an actual expert. In both cases, the procedure is identical to the previous method, with the difference that instead of using various distinguishing sequences, we use a single checking sequence, and the expected output is given by the oracle or the expert. The stop condition is the equality in the checking sequence's output and the output given by the expert or the oracle.    The repair process can also stop if  a maximal number of repair attempts is reach. Results show that this repair method can not prevent the re-generation of already visited DFSMs. The advantage of using checking sequences for repairing generated DFSM, is that only one sequence needs to be analysed by a human expert for building the repair query. 
Algorithm \ref{alg:repair:checkseq} presents the repair procedure. 
\begin{algorithm}[!t]
\caption{Repair procedure based on Checking Sequence}\label{alg:repair:checkseq}
\begin{algorithmic}
\Require A description $d$ of a DFSM
\Require $n$ a maximal number of states
\Require $K$ the maximal number of repair attempts. 
\State build the query $\pi$ for $d$
\State Set $\pi_r= \pi$
\While{$K > 0$}
    \State Generate a DFSM $\mathcal{S}$ by querying GPT-4.0 with $\pi_r$
    \State Transform $\mathcal{S}$ into an input complete DFSM, if $\mathcal{S}$ is not input complete  
    \State Estimate the checking sequence $\sigma/\gamma$ of $\mathcal{S}$ 
    \State Get the expected output $\gamma_{exp}$ for input sequence $\sigma$ from an expert
    \If {$\gamma==\gamma_{exp}$} 
    \State Return "$\mathcal{S}$ is the expected DFSM."
    \Else
        \State Build a repair query $\pi_\tau$ for $\tau = \sigma/\gamma_{exp}$
        \State Set $\pi_r = \pi_r\pi_\tau$
        \State Set $K= K-1$
    \EndIf
\EndWhile
    \State Return "repair failure"
\end{algorithmic}
\end{algorithm}

Table~\ref{table:approach:checkseq:result} presents the evaluation results  of the effectiveness of repair, out of the 5 faulty generated machines 1 was repaired in 1 attempt, 1 in 3 attempts and 3 were not repaired. 
\begin{table}[t]
    \centering
\begin{tabular}{|c|c|p{2cm}|p{2cm}|} \hline 
   {\bf nb states}  & nb of oracles & nb. of  faulty generated DFSM & repair succeeding rate \\ \hline  
   5 & 30 & 5 & 40\% \\ \hline 
\end{tabular}
    \caption{Evaluation results of the repair approach based on checking sequences}
    \label{table:approach:checkseq:result}
\end{table}
Results in Table~\ref{table:approach:checkseq:result} consider DFSM having five states. This is because computing checking sequences is time-consuming, even for machines with limited number of states (e.g., 8, 9, etc.). The results indicate that DFSM repair based on refining prompt with checking sequences may not be efficient or require devising more efficient prompts. 

\subsection{Approach based on Fault models}
Estimating  checking sequences is complex, and checking sequences are often very long, which could hamper the application of the previous approach.   Prior knowledge about the weaknesses of GPT-4 allows us to identify the most common faults~\cite{nikbin2024analysis} introduced in the produced specifications. This knowledge can be used to devise a  repair domain of DFSM repair candidates and to elaborate an efficient method for mining a proper candidate from this  repair domain. 

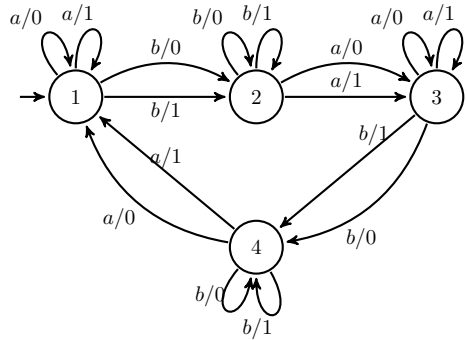
\begin{figure}[t]
    \centering
    \scalebox{0.80}{
\begin{tikzpicture}[->,>=stealth',shorten >=1pt,auto,node distance=2.5cm,initial text={},
every initial by arrow/.style={->}]
{\fontsize{10}{10}\selectfont
\tikzstyle{every node}=[scale=1,inner sep=1pt, outer sep=1pt,line width=1pt]
\tikzstyle{every edge}=[scale=1,line width=1pt, draw]
\tikzstyle{every state}=[fill=white,draw,text=black]
  \node[initial,state]  (S1)   at (0,5)                {$1$};
  \node[state]         (S2)  at ([xshift=3cm,yshift=0cm]S1) {$2$};
  \node[state]         (S3) at ([xshift=3cm,yshift=0cm]S2) {$3$};
  \node[state]         (S4) [below of=S2] {$4$};
  \path (S1) edge[loop, in=60, out=90,looseness=10]       node[xshift=-0.6cm]  {$a/1$} (S1)
             edge [loop, in=100, out=130,looseness=10] node {$a/0$} (S1)
             edge[bend left]          node[] {$b/0$}(S2)
             edge[]          node[below] {$b/1$}(S2)
             (S2)  edge[loop, in=60, out=90,looseness=10]       node[xshift=-0.6cm] {$b/1$}(S2)
             edge [loop, in=100, out=130,looseness=10] node {$b/0$} (S2)
             edge[bend left]  node[above, pos=0.5] {$a/0$} (S3)
             edge[]  node[above, pos=0.5] {$a/1$} (S3)
             (S3) edge [loop, in=100, out=130,looseness=10] node {$a/0$ } (S3) 
             edge [loop, in=60, out=90,looseness=10]       node[xshift=-0.6cm] {$a/1$ } (S3)
             edge[bend left] node[pos=0.7] {$b/0$ } (S4)
             edge[] node[pos=0.3,above] {$b/1$ } (S4)
             (S4) edge [loop , in=-100, out=-130,looseness=10]  node[xshift=-0.6cm] {$b/0$ } (S4) 
             edge [loop , in=-90, out=-60,looseness=10]  node[xshift=-0.6cm]  {$b/1$ } (S4)
             edge[above] node {$a/1$ } (S1)
             edge[bend left] node {$a/0$ } (S1) 
             ;
}
\end{tikzpicture}
}
    \caption{Illustration of a mutation machine defining a repair domain consisting of $256$ DFSMs, one of them can be the expected DFSM. This machine is built from the DFSM in~Figure~\ref{fig:fsm:running:graph} by applying mutation operations corresponding to only output faults only. }
   \label{fig:mut:machine}
\end{figure}

\emph{Mutation machines}~\cite{koufareva1999test} can serve to represent the repair domain. A mutation machine is nondeterministic finite state machine (FSM) that includes (defines) a huge number of DFSM specifications. It can be built based on knowledge  of types of faults LLMs can introduce in generated DFSMs. Each type of faults serve to add new transitions in the generated DFSMs. The new transitions corresponds to plausible patches for repairing the generated DFSMs.   Figure~\ref{fig:mut:machine} presents a mutation machine for the DFSM produced from Description in Listing{lst}. The machine adds patches (transitions) in state $3$ for input $b$ and input $a$. It also add a patch in state $4$ on input $a$.   The repair procedure aims at mining the correct DFSM from the repair domain, which is equivalent to choose the adequate patches.   An algorithm for mining algorithm is iterative. Each iteration step determines a distinguishing sequence for two DFSM $\mathcal{S}_1$ and $\mathcal{S}_2$ included in the fault domain and a list of output sequences that can be produced by all DFSMs included in the fault domain. The algorithm queries an expert for obtaining the expected output sequence which is then used to identify and remove incorrect DFSM  from the fault domain. We consider SAT-based mining algorithm in~\cite{nguena2021mining} and DFSM oracles playing the role of human experts. This repair procedure always terminates. It produce the repaired DFSMs if it is included in the repair domain; otherwise a repair failure occurs.  

Table~\ref{table:approach:fault-model:result} presents the evaluation results of the fault model-based repair approach. We consider a new prompt design for demonstrating the robustness of the whole design  approach against the prompt engineering.  The mutation machines are built from  generated DFSMs by assuming   that GPT-4.0 introduces output faults and missing transition faults, which we observed. We built the repair domains based on these observations. Eventually we add specific transitions, when it is needed,  for the repairs domain to include the oracles. This ensure that the repair procedure will always converge in our experiments. Notably, the addition of specific transitions occurred only a few times during our experiments, which suggests that the automatic construction of the repair domain, informed by an understanding of LLM weaknesses, is both feasible and effective. 
In practice, oracles are often unavailable, making it impossible to add specific transitions. In this scenario, the repair procedure generates a DFSM within the repair domain that aligns with all expert-provided answers to output queries, if such a DFSM exists. The generated DFSM can subsequently be validated using model-based testing techniques.

Experimental results shows experts could answer to an important number of output queries in the repair process. Method are needed to reduce this number. A plausible solution could consider generating a single checking sequences for identifying DFSM in  reduced repair domains. We foresee that such checking sequences will be shorter. 

\begin{table}[t]
\centering
    \begin{tabular}{|p{4cm}|c|c|} \hline 
   {\bf nb states}  & 5 & 10 \\ \hline 
   { nb. of auto. gen. oracles} & 30 & 30 \\ \hline 
   { nb. of  faulty generated DFSM} & 15 &  26 \\ \hline 
   { repair succeeding rate} &100\% & 100\% \\ \hline
   { max nb. of output queries}& 30& 89  \\ \hline 
   { max length of output queries} & 4 & 5 \\ \hline 
   nb. repair domains augmented with specific transitions & 4 & 12 \\ \hline
   { nb. max of added specific transitions} & 4 & 5\\\hline 
\end{tabular}
    \caption{Evaluation results of the repair approach based on fault model}
    \label{table:approach:fault-model:result}
\end{table}

\section{Conclusion}
We have proposed an approach for designing formal specifications of reactive systems from textual requirements using the LLM GPT-4. This approach involves prompting the LLM to generate DFSM specifications from descriptions and repairing any faults in the generated specifications.
We used randomly generated descriptions to empirically evaluate the precision of this approach. The results showed that some generated specifications conceal faults.\\ 
We detected faults in the specification with methods based the syntax and semantics of the specification. We elaborated four repair approaches for the generated specifications. First, the syntax-based approach, though unrealistic, demonstrates that refining the prompt with identified syntactic faults improves the LLM's attention to its previous errors and allows quasi-full repair of faulty specifications. 
In contrast, the first semantics-based approach, which involves determining distinguishing sequences to enhance the prompts, yielded significantly different results. This theoretical and not realistic method, requiring a much more precise understanding and analysis of the specified behavior, has served to demonstrate that is it unlikely to repair generated specifications by refining the prompt with numerous input-output sequences representing the (un)expected behaviors of the DFSMs. 
Our experiments indicate that either GPT-4 struggles with this level of analysis, or more advanced prompts are necessary." \\
The Second semantics-based approach is semi-automatic and much more realistic and applicable. It utilises automatically constructed checking sequences and expert knowledge of the system under design to determine the expected output of a unique checking sequence. This information is used to refine the LLM's prompt. 
The results using this method are slightly worst than the first semantic-based method, especially because the computation of checking sequences is time-consuming. Moreover a single and long checking sequence simultaneously gathers information about  multiple parts of the branching structure of DFSMs at the same time. Utilising specialised attention mechanisms and more sophisticated prompts could further enhance the effectiveness of this approach. \\ 
The third semantics-based approach leverages a better understanding of the faults introduced by the LLM. It uses a repair domain to delimit  possible repair spaces for the generated specifications and involves interactions with experts to determine output sequence corresponding to automatically generated relevant input  sequences allowing for the automatic selection of  adequate repairs. This approach does not involve LLM prompting for the repair, instead it relies on formal methods of  mining specifications from  mastered repair domains. The approach is promising and requires a deeper understanding of the faults that LLMs can make, which will be the subject of future work. \\
We also observed that as the size of the machine increases, the volume of information provided to the LLM also grows, leading to a corresponding rise in the number of faults produced by the LLM.  LLM seems to lack the capacity to perform sharp analysis, and the capacity to have an "intelligent" understanding of connections (input-output sequences and transitions to correct) and suffer from a lack of precision when handling higher amounts of information. Nevertheless, to realistic scales, LLMs perform well when generating DFSM from informal specifications, and with a lesser intervention of LLMs, tangible and realistic repair methods can be produced, such as the repair based on fault models.\\
Additionally, future work considers applying this approach to descriptions of industrial-like systems. We foresee, based on our preliminary experiments with examples in~\cite{steffen2011introduction}, that the industrial-like description will add domain-specific information permitting LLM to generate correct specifications or repair the generated specification with the repair prompts built from input-output sequences. 

\bibliographystyle{splncs04}
  \bibliography{biblio}

\newpage 

\section*{Appendix }
\begin{lstlisting}[caption=Patterns for  generating textual description of state, label=listing:pattern:state]
def toNL(self) :
  descCandidate = [f"state {self._label}",f"{self._label}"]
  rst = random.choice(descCandidate)
  return rst
\end{lstlisting}

\begin{lstlisting}[ caption=Patterns for  generating textual description of transitions, label=listing:pattern:transition]
def toNL(self) -> str :
   liste = [f"{self._fromtoNL()} {self._outputtoNL()} and {self._movetoNL()} {self._inputtoNL()}", 
            f"{self._outputtoNL()} and {self._movetoNL()} {self._inputtoNL()} {self._fromtoNL()}"]
   rst = random.choice(liste)
   return rst

def _systemtoNL(self) ->str :
   liste = ["it", "the system", "the application"]
   rst = random.choice(liste)
   return rst

def _fromtoNL(self) -> str :
liste = [f"from {self._src.toNL()}",
         f"from state {self._src.toNL()}",
         f"in state {self._src.toNL()}",
         f"in  {self._src.toNL()}",
         f"when the system is in {self._src.toNL()}",
         f"when it is in {self._src.toNL()}"]
    rst = random.choice(liste)
    return rst

def _movetoNL(self) -> str :
    liste = [f"{self._systemtoNL()} moves to {self._tgt.toNL()}",
             f"{self._systemtoNL()} reaches {self._tgt.toNL()}",
             f"{self._tgt.toNL()} is reached"]
    rst = random.choice(liste)
return rst

def _inputtoNL(self) -> str :
    liste= [f"if the input is {self._input}",
            f"if the input {self._input} occurs",
            f"if  {self._input} occurs",
            f"on occurence of input {self._input}",
            f"on occurence of {self._input}"]
    rst = random.choice(liste)
    return rst

def _outputtoNL(self) -> str :
    liste= [f"{self._systemtoNL()} produces {self._output}",
            f"{self._systemtoNL()} returns {self._output}",
            f", {self._output} is produced",
            f"the output {self._output} is produced",
            f", {self._output} is returned"]
    rst = random.choice(liste)
    return rst
\end{lstlisting}

\begin{lstlisting}[ caption=Patterns for  generating textual description of FSM, label=listing:pattern:fsm]
def toNL(self) -> str :
  rst =""
  for cle in self._transitionsById.keys() :
     rst +=  " "+self._transitionsById[cle].toNL() + ".\n"
  return rst 
\end{lstlisting}

\end{document}